\documentstyle[11pt]{article}
\def\fnote#1#2{\begingroup\def\thefootnote{#1}\footnote{#2}\addtocounter
{footnote}{-1}\endgroup}

\begin{document}

\hfill{UTTG-10-16}

\vspace{36pt}

\begin{center}
{\large {\bf {Lindblad Decoherence in Atomic Clocks}}}

\vspace{36pt}
Steven Weinberg\fnote{*}{Electronic address:
weinberg@physics.utexas.edu}\\
{\em Theory Group, Department of Physics, University of
Texas\\
Austin, TX, 78712}

\vspace{30pt}

\noindent
{\bf Abstract}
\end{center}

\noindent
It is shown how possible corrections to ordinary quantum mechanics described by the Lindblad equation might be detected by exploiting the great precision of atomic clocks.  

\vfill

\pagebreak

In searching for an interpretation of quantum mechanics we seem to be faced with nothing but bad choices[1].  
To avoid both the dualism of the Copenhagen interpretation and   the endless creation of inconceivably many branches of history of the many-worlds approach, while at the same time holding on to a realist description of the evolution of physical states from moment to moment,  we may try to modify quantum mechanics so that during measurement the density matrix of even an isolated system undergoes a   collapse of the sort called for by the Copenhagen interpretation.  The idea is that this collapse is rapid in systems containing macroscopic elements, such as apparatus or physicists, while the corrections to quantum mechanics are very small in purely microscopic systems such as atoms.    A collapse of the density matrix has already been described theoretically in interesting modifications of quantum mechanics[2].  Here we wish to explore the possibility of observing small departures from quantum mechanics by exploiting the  great precision of atomic clocks.  Apart from this aim, the formalism developed here may prove useful in describing limits on the precision of atomic clocks in ordinary quantum mechanics due to their interaction with the environment.

First, a reminder of the time-dependence to be expected   both in modified versions of quantum mechanics and in open systems.  To avoid instantaneous communication at a distance[3], the density matrix at time $t'$ is assumed to  depend only on the density matrix at any earlier time $t$, but not otherwise on the state vector at earlier times.  Following the rules for composition of probabilities, we take this to be a  linear relation.   We require that this relation preserves the trace and the Hermiticity of the density matrix, and satisfies a condition of complete positivity[4].
It is well-known  that under these assumptions the time-dependence of the density matrix is given by a  first-order differential equation,  the Lindblad equation[5]:
\begin{equation}
\dot{\rho}(t)=-i[H,\rho(t)]+\sum_\alpha \Bigg[L_\alpha\,\rho(t)\,L_\alpha^\dagger
-\frac{1}{2}L^\dagger_\alpha\,L_\alpha\,\rho(t)
-\frac{1}{2}\rho(t)\,L_\alpha^\dagger\,L_\alpha\Bigg]\;.
\end{equation}
We are here considering only a Hilbert space of finite dimensionality $d$, which is adequate for  the application we have in mind.  (We ignore the translational degree of freedom of atoms.)  In Eq.~(1), 
 $H$ is a $d\times d$ Hermitian matrix that can be identified with the Hamiltonian of ordinary quantum mechanics, and  the sum   runs over  not more than $ d^2-1$ matrices $L_\alpha$, which represent the departure from ordinary quantum mechanics.  We use units with $\hbar=1$.

The form of Eq.~(1)  also assumes time-translation invariance, in which case the relation between $\rho(t)$ and $\rho(t')$ depends only on $t'-t$, and the matrices $H$ and 
$L_\alpha$ are time-independent.  This is of course not the case throughout the history of  atoms in an atomic clock, which are intermittently exposed to external electromagnetic radiation.  But  atomic clocks rely on a ``Ramsey trick''[6], in which atoms are exposed to electromagnetic radiation only in two relatively short bursts, separated by a much longer interval in which they are free of external fields.  It is this long time interval between bursts that gives the atomic wave function a chance to get out of phase with the electromagnetic wave,  and so leads to the high precision with which the frequency of the wave can be tuned to that of the atomic transition, and it is also during this field-free period that the small effects due to the corrections to ordinary quantum mechanics have a chance to build up.  So to deal with atomic clocks we shall first  consider the field-free case, with time-dependence prescribed by the time-independent equation (1), and then return to the clocks. 
  
We will simplify our task here by assuming (in agreement with observation) that the states $|m\rangle$ with which we have to deal are stable, aside from radiative transitions that are slow enough to be ignored.   
We shall also assume that Eq.~(1) does not allow a decrease in the von Neumann entropy $-{\rm Tr}\Big(\rho\,\ln\rho\Big)$ for any $\rho$.  It follows then that the stable states are eigenstates of $L_\alpha$, $L_\alpha^\dagger$, and $H$, with eigenvalues that we shall call $\ell_{\alpha m}$, $\ell^*_{\alpha m}$, and $E_m$.  

Here is the proof[7].  If $|m\rangle$ is stable then  the right-hand side of Eq.~(1) must vanish if we take $\rho$ to be the projection operator $\Lambda_m=|m\rangle\langle m|$ on such a state.  Multiplying this equation on the left with $\Lambda_m$ and taking the trace, with a little rearrangement we have
\begin{equation}
0={\rm Tr}\left\{\sum_\alpha [L_\alpha,\Lambda_m]^\dagger[L_\alpha,\Lambda_m]\right\} 
+{\rm Tr}\left\{\Lambda_m\sum_\alpha \Big(L_\alpha^\dagger L_\alpha-L_\alpha L_\alpha^\dagger\Big)\right\}\;.
\end{equation}
The necessary and sufficient condition for the non-decrease of entropy is the vanishing of the sum over $\alpha$ in the second term[8], leaving us here with the vanishing of the first term, and hence with the vanishing of $[L_\alpha,\Lambda_m]$ for all $\alpha$.  The adjoint shows that also $[L^\dagger_\alpha,\Lambda_m]$ vanishes for all $\alpha$.  Then the vanishing of the right-hand side of Eq.~(1) where $\rho=\Lambda_m$ requires also that $[H,\Lambda_m]=0$.  Letting these vanishing commutators act on $|m\rangle$ shows 
immediately that $|m\rangle$ is an eigenstate of $L_\alpha$, $L_\alpha^\dagger$, and $H$, as was to be proved.

Note that we have not had to assume that the discrete stable states form a complete set.  Indeed, we only need to assume stability for the two states involved in the clock transition.  To digress a bit, if we had assumed that the stable states form a complete set,  then we could have concluded from the above that these states would form a basis in which $L_\alpha$, $L^\dagger_\alpha$, and $H$ are all diagonal, so that they would all commute with each other, and so energy would be conserved --- not a surprising conclusion, thought it could not have been reached here without our further assumption of non-decreasing entropy.  The conservation of energy by the Lindblad equation might raise problems with locality and Lorentz invariance[9], though this is uncertain[10].

In accordance with this theorem,  Eq.~(1) gives the density matrix the time-dependence
\begin{equation}
\rho_{mn}(t)\propto \exp \Big[-i(E_m-E_n)t-\lambda_{mn}t\Big]
\end{equation}
where $|m\rangle$ and $|n\rangle$ are any two stable states, and 
\begin{eqnarray}
&&\lambda_{mn}=\sum_\alpha \Bigg[\frac{1}{2}\Big|\ell_{\alpha m}\Big|^2+\frac{1}{2}\Big|\ell_{\alpha n}\Big|^2-\ell_{\alpha m}\ell_{\alpha n}^*\Bigg]
\nonumber\\&&
=
\sum_\alpha \Bigg[-i\,{\rm Im}(\ell_{\alpha m}\ell_{\alpha n}^*)+\frac{1}{2}\Big|\ell_{\alpha m}-\ell_{\alpha n}\Big|^2 \Bigg]\;.
\end{eqnarray}
We note that ${\rm Re}\lambda_{mn}\geq 0$, so all elements of the density matrix decay except for those with ${\rm Re}\lambda_{mn}=0$,  Also, $\lambda_{mm}=0$, 
so the diagonal elements  $\rho_{mm}(t)$, and typically only the diagonal elements, are time-independent. 

Now let us see what this implies for the tuning of the frequency of an electromagnetic wave to the transition frequency $E_e-E_g$ between stable states $|g\rangle$ and $|e\rangle$ in an atomic clock.  (The labels $e$ and $g$ are conventional, standing for ``excited state'' and ``ground state,'' though $g$ and $e$ can be any two stable states of the atom.)
Each atom is exposed twice for periods each lasting  a relatively short time $\tau$ to an oscillating external electromagnetic field, which adds to the Hamiltonian a term $H'\exp(-i\omega t)+H'^\dagger\exp(i\omega t)$, and can drive the transition $g\rightarrow e$ when the real frequency  $\omega$ is tuned to a value near $E_e-E_g$.    We will work with an ``interaction picture'' density matrix $\rho^I_{mn}(t)\equiv \exp(i(E_m-E_n)t)\rho_{mn}(t)$.  We assume that the exposure period $\tau$ is short enough so that $\tau|\lambda_{mn}|\ll 1$, and hence during this period changes in $\rho^I$ arise only from the oscillating external field.  We make the usual assumptions that   $\tau |E_e-E_m|\gg 1$ for $m\neq e$, $\tau |E_g-E_m|\gg 1$ for $m\neq g$   and $\tau|\omega|\gg 1$, which allows us to drop rapidly oscillating terms in the equation for $\dot{\rho}^I$ and keep only those terms with time-dependence proportional to $\exp(\pm i\Delta\omega t)$, where $\Delta\omega\equiv \omega-E_e+E_g$.  We also suppose  that as usual in atomic clocks the frequency of the external field has been tuned so that $|\Delta\omega|\ll |H'_{eg}|$, and hence the frequency of Rabi oscillations is $\Omega/2=|H'_{eg}|$.  Under these assumptions, the density matrices at times $t$ and $t+\tau$ are related  by 
\begin{equation}
\rho^I(t+\tau)=U(t+\tau, t)\rho^I(t)U^\dagger(t+\tau, t)\;,
\end{equation}
where 
\begin{eqnarray}
&& U_{ee}(t+\tau,t)=U_{gg}(t+\tau,t)=\cos(\Omega\tau/2)\;,~~~~~\;,\nonumber\\
&& U_{eg}(t+\tau,t)=U_{ge}^*(t+\tau,t)=-ie^{i\Delta\omega\, t}\sin(\Omega \tau/2)\;,
\end{eqnarray}
(We are choosing the relative phase of the states $e$ and $g$ so that $H'_{eg}$ is real and positive, and hence equal to $\Omega/2$.)

If an atom starts at $t=0$ in the pure state $g$, then 
at  time $t=\tau$ its density matrix $\rho^I(t)= U(\tau, 0)\rho^I(0)U^\dagger(\tau, 0)$ will have components
\begin{eqnarray}
&& \rho_{ee}^I(\tau)=\sin^2(\Omega\tau/2)\;,~~~~~\rho_{gg}^I(\tau)=\cos^2(\Omega\tau/2)\;,\nonumber\\
&& \rho_{eg}^I(\tau)=\rho_{ge}^{I*}(\tau)=ie^{-i\Delta\omega\tau}\cos(\Omega \tau/2)\;\sin(\Omega\tau/2)\;,
\end{eqnarray}
which of course still represents a pure state.
  
Then for a Ramsey time $T\gg \tau$ the atom travels through field-free space, so the only time-dependence of the density matrix $\rho^I$ in this period arises from the Lindblad term in Eq.~(1).  In accordance with Eq.~(3), the density matrix at the end of this period is
\begin{eqnarray}
&& \rho_{ee}^I(\tau+T)=\sin^2(\Omega\tau/2)\;,~~~~~\rho_{gg}^I(\tau)=\cos^2(\Omega\tau/2)\;,\nonumber\\
&& \rho_{eg}^I(\tau+T)=\rho_{ge}^{I*}(\tau+T)=ie^{-i\Delta\omega\tau}e^{-\lambda_{eg}T}\cos(\Omega \tau/2)\;\sin(\Omega\tau/2)\;.
\end{eqnarray}

Then in  a second period of duration $\tau$ the atom is again exposed to the same external electromagnetic field, and the density matrix is changed to
\begin{equation}
\rho^I(2\tau+T)=U(2\tau+T,\tau+T)\rho^I(\tau+T)U^\dagger(2\tau+T,\tau+T)
\end{equation}
A straightforward calculation gives the probability $P_e$ that the atom will wind 
up in the excited state:
\begin{equation}
P_e=\rho^I_{ee}(2\tau+T)=\frac{1}{2}\sin^2\Omega\tau\left[
1+e^{-\Gamma T}\cos\Bigg(\Big(\omega-E_e+E_g-{\cal E}\Big)T\Bigg)\right]\;,
\end{equation}
where we write $\lambda_{ge}=\Gamma-i{\cal E}$ with $\Gamma$ and ${\cal E}$ real, and hence
according to Eq.~(4),
\begin{equation}
\Gamma=\frac{1}{2}\sum_\alpha\Big|\ell_{\alpha g}-\ell_{\alpha e}\Big|^2 \Bigg] 
\end{equation}
\begin{equation}
{\cal E}=-\sum_\alpha{\rm Im}\left\{\ell_{\alpha g}\ell^*_{\alpha e}\right\}
\end{equation}
In using atomic clocks the excitation probability $P_e$ is measured as a function of frequency $\omega$ by repeating the observation of the fraction of atoms excited for various chosen frequencies $\omega$.  Then $\omega$  is tuned to maximize $P_e$, so that $\omega$ will  then normally be expected to equal the reference frequency  $E_e-E_g$  within an uncertainty of order $1/T$.  

If there were corrections to ordinary quantum mechanics in Eq.~(10) with $\Gamma$ of order $1/T$ or greater, the shape of  the curve of $P_e$ versus $\omega$ would be grossly altered.  For instance, for $\Gamma T=1$, the ratio of the minimum value of $P_e$ to its maximum value would be 0.46 instead of zero, and the ratio of the value of $P_e$ where it is most rapidly varying with frequency to its maximum value would be 0.73 instead of 0.5.  Seeing such a departure from expectations would be a good sign of a departure from ordinary quantum mechanics.  A change in the form of $P_e$ versus $\omega$ this drastic would generally have been seen in atomic clocks and has not been seen[11], so it seems safe to conclude that $\Gamma$ is less than the  values of $1/T$ encountered in atomic clocks.

Unfortunately we have no idea of what target value of $\Gamma$ which we should aim at, or even how $\Gamma$ might vary from one transition to another.  We can distinguish two extreme cases.

    If $\Gamma$ has similar values for all transitions, then we should look at clocks for which the Ramsey time $T$ is as long as possible.  Modern atomic clocks typically have $T$ of the order of seconds, but a clock[12] using  a microwave-frequency transition in trapped ${}^{171}{\rm Yb}^+$ ions has operated with $T> $ 600 seconds.  Hence we can conclude that  
in this transition $\Gamma< 10^{-18}$ eV.  This upper limit shows that environmental effects make it   hopeless to look for departures from quantum mechanics on macroscopic scales, where the energy of interaction with the environment is presumably always much greater than $ 10^{-18}$ eV.  On the other hand, this upper bound is enormous compared with the difference between energies of discrete states of macroscopic objects that are free from all external influences.  For instance, according to quantum mechanics, the successive energy eigenstates of a pointer of mass one gram and length one centimeter that swivels freely in two dimensions is about $10^{-42}$ eV.  Thus departures from ordinary quantum mechanics with $\Gamma$ less than the limit $10^{-18}$ eV derived from atomic clocks might still have a powerful effect on the quantum states of macroscopic systems if they could somehow be isolated from their environment.  

If instead $\Gamma$  somehow scaled with the transition frequency $E_e-E_g$, then we would want to set a limit on $\Gamma/(E_e-E_g)$, rather than on $\Gamma$ itself.  For this purpose it would be more useful to look at clocks for which 
the fractional imprecision $1/T(E_e-E_g)$ is as small as possible.   For optical clocks with $T$ of the order of a second this is $10^{-15}$, but a clock using ${}^{37}{\rm Al}$ ions achieved a value about $3\times 10^{-17}$[13], so we can conclude that at least for these transitions, $\Gamma/(E_e-E_g)< 3 \times 10^{-17}$.

In addition to a change in the shape of the curve of $P_e$ versus $\omega$, Eq.~(10) also entails a shift in the frequency at the maximum value of 
$P_e$, from $E_e-E_g$ to $E_e-E_g+{\cal E}$.  
Detecting this frequency shift   is impossible in a two-state system if we do not have independent information about the uncorrected frequency $E_e-E_g$.  The prospects are brighter if it is possible to drive transitions among three different energy levels, because actual energy differences trivially obey the relation
$$ (E_1-E_2)+(E_2-E_3)+(E_3-E_1)=0$$
while there is no reason to expect the frequency shifts ${\cal E}_{ij}$ to obey the corresponding relation 
$${\cal E}_{12}+{\cal E}_{23}+{\cal E}_{31}=0\;.$$ 
It remains to be seen if there is a three-level system suitable for this purpose.

\vspace{20pt}

I am grateful for helpful conversations about atomic clocks with Mark Raizen and David Wineland.  This material is based upon work supported by the National Science Foundation under Grant Number PHY-1620610 and with support from The Robert A. Welch Foundation, Grant No. F-0014.

\vspace{10pt}

\begin{center}
{\bf ---------}
\end{center}

\vspace{10pt}

\begin{enumerate}

\item The author's views on this issue are set out in detail in Section 3.7 of S. Weinberg, {\em Lectures on Quantum Mechanics}, 2nd ed. (Cambridge University Press, Cambridge, U. K., 2015).
\item G. C. Ghirardi, A. Rimini,
and T. Weber, Phys. Rev. D {\bf 34}, 470 (1986);  P. Pearle, Phys. Rev. A {\bf 39}, 2277 (1989), and in {\em Quantum Theory: A Two-Time Success Story} (Yakir Aharonov Festschrift), eds. D. C. Struppa \& J. M. Tollakson (Springer, 2013), Chapter 9. [arXiv:1209.5082]; G. J. Milburn, Phys. Rev. A {\bf 44}, 5401 (1991); I. C. Percival, Proc. Roy. Soc. London A, {\bf 447}, 189 (1994).  For a review, see A. Bassi and G. C. Ghirardi, Physics Reports {\bf 379}, 257 (2003).  
\item N. Gisin, Helv. Phys. Acta {\bf 62}, 363 (1989); Phys. Lett. A {\bf 143}, 1 (1990).  This is discussed in a wider context by J. Polchinski, Phys. Rev. Lett. {\bf 66}, 397 (1991).
\item W. F. Stinespring, Proc. Am. Math. Soc. {\bf 6}, 211 (1955).  For a review, see F. Benatti and R. Florentini,  Int.J.Mod.Phys. {\bf B19}, 3063 (2005) [arXiv:quant-ph/0507271].
\item G. Lindblad, Commun. Math. Phys. {\bf 48}, 119 (1976); V. Gorini, A. Kossakowski and E. C. G. Sudarshan, J. Math. Phys. {\bf 17}, 821 (1976).  For a lucid derivation of the Lindblad equation see P. Pearle, arXiv: 1204.2016.
\item N. F. Ramsey, Phys. Rev. {\bf 76}, 996 (1949).
\item The proof follows along the same lines as in S. Weinberg, Phys. Rev. A {\bf 93}, 032124 (2016).
\item F. Benatti and R. Narnhofer, Lett. Math. Phys. {\bf 15}, 325 (1988).
 \item T. Banks, M. E. Peskin, and L. Susskind, Nucl. Phys. {\bf B244}, 125 (1984); M. Srednicki, Nucl. Phys. {\bf B410}, 143 (1993).  
\item  W. G. Unruh and R. M. Wald, Phys. Rev. D {\bf 52}, 2176 (1995).    
\item D. J. Wineland, private communication.
\item P. T. H. Fink, M. J. Sellars, M. A. Lawn, C. Coles, A. G. Mann, and D. G. Blair, IEEE Transactions on Instrumentation and Measurement {\bf 44}. 113 (1995).
\item C. W. Chou, D. B. Hume, M. J. Thorpe, D. J. Wineland, and T. Rosenband, Phys. Rev. Lett. {\bf 106}, 160801 (2011).
\end{enumerate}

  \end{document}